\documentclass[twocolumn,preprintnumbers,amsmath,amssymb,pra]{revtex4}

\usepackage{graphicx,amsmath}
\usepackage{dcolumn}
\usepackage{bm}
\usepackage{amssymb}
\usepackage{epstopdf}
\usepackage{color}

\begin{document}

\title{Waveguide bandgap N- qubit array with a tunable transparency resonance}

\begin{abstract}
We study a single photon transmission through 1D N- qubit chain.
The qubits are supposed to be identical with equal distance
between neighbors. We express the transfer matrix of N- qubit
chain in terms of Chebyshev polynomials, which allows us to obtain
simple expressions for the transmission  and reflection amplitudes
for arbitrarily large N. If the distance between neighbor qubits
is equal to half wavelength, the transmission spectrum exhibits a
flat bandgap structure with very steep walls. We show that for odd
N the tuning of the excitation frequency of a central qubit gives
rise to the appearance within a bandgap of a narrow resonance with
a full transmission. The position of the resonance and its width
can be controlled by the frequency of a central qubit. We show
that the formation of the bandgap and of the transmission
resonance is conditioned by the overlapping the widths of
individual qubits which results from the strong coupling between
qubits and waveguide photons.

\end{abstract}


\date{\today }

\author{Ya. S. Greenberg}\email{yakovgreenberg@yahoo.com}
\affiliation{Novosibirsk State Technical University, Novosibirsk,
Russia}
\author{A. A. Shtygashev} \affiliation{Novosibirsk State
Technical University, Novosibirsk, Russia}
\author{A. G. Moiseev} \affiliation{Novosibirsk State
Technical University, Novosibirsk, Russia}


 \maketitle

\section{Introduction}\label{intr}
One-dimensional (1D) waveguide-quantum electrodynamics (QED)
systems are promising candidates for quantum information
processing \cite{Roy2017}. One important implementation is
waveguide QED, where a quantum multi- qubit system coupled to a
common waveguide interacts coherently with the continuum of modes
of a waveguide instead of a cavity\cite{Loo2013, Lalum2013,
Alb2019, Zhang2019, Greenberg15}.


 The
multiple qubit system obtains an infinite range photon mediated
effective interaction which can be tuned with the inter-qubit
distance. Furthermore, this system exhibits collective excitations
with lifetimes ranging from extremely sub- to superradiant
relative to the radiative lifetime of the individual qubits
\cite{Alb2019, Zhang2019, Brehm2020, Mir2019}.

Among the qubit family, the chain which consists of $N$ identical
equally spaced qubits is the most simple one. There are many
papers where the photon transmission through this structure has
been analytically treated, however, here we refer only to the
papers  \cite{Tsoi2008, Mukh2019} where the analytical treatment
is similar to the one used in this paper.

In the present paper we propose a matrix formalism for the
calculation of a transmission amplitude in 1D open waveguide
filled with a number of equally spaced identical qubits. The
formalism is based on the decomposition of the transfer matrix for
N- qubit chain in terms of Chebyshev polynomials. This method has
originally been developed for the description of light scattering
in periodic stratified media \cite{Abel1950, Yeh1977}.

In Section II we express transfer matrix for equally spaced $N$
identical qubits in terms of Chebyshev polynomials, which allows a
concise expression for transmission and reflection amplitudes. We
show that if the distance $L$ between neighbor qubits is equal to
a half wavelength at the qubit frequency $\Omega$, the
transmission spectrum exhibits a bandgap with steep walls. The
width of the bandgap weakly depends on the qubit number $N$,
whereas the suppression of transmitted signal at the bandgap
boundaries scales as $N^{-2}$. The existence of this bandgap is
not surprising because it is a common feature of many periodic
structures \cite{Yeh1977, Bend1996, Deutsch1995, Mirza2017,
Ruos2017, Ivch1994, Ivch2013}. In principle, bandgaps appear at
any nonzero value of the qubit phase $k_0L=\Omega L/v_g$, where
$v_g$ is the group velocity \cite{Mukh2019}. However, the value
$k_0L=\pi$ has a unique property, which is studied in Section III.

In Section III we consider the $N$- odd qubit chain where $N-1$
qubits have the same excitation frequency $\Omega$, whereas the
excitation frequency $\Omega_0$ of a central qubit is different.
In general, if we add new energy $\Omega_0$ to the system it can
give rise to new resonance, as was experimentally shown in
\cite{Brehm2020}. However, we show that for our structure there
exists inside the gap a frequency where the reflection amplitude
equals exactly zero. Therefore, within the bandgap a narrow
resonance line appears which provides a full transmission. The
position of this resonance and its width depend on the ratio
$\Omega_0/\Omega$. Therefore, in our case the resonance line which
gives rise to a full transmission, moves within the bandgap with
its position and the width being controlled by the excitation
frequency of the central qubit.

We show that both the bandgap and the narrow resonance result from
strong photon- qubit coupling when the width of individual qubits
become overlapped. From the quantum point of view they represent
superradiant and subradiant states, respectively.

At the end of this section we consider how the radiation losses to
the environment influence the photon transmission trough this
structure.

The basic properties of Chebyshev polynomials, which are used in
the main text are given in the Appendix.

\section{Transfer matrix for N identical qubits}
Our approach is based on the  transfer matrix for a single photon
scattering on two-level atom or qubit which is written as follows
\cite{Shen05}:

\begin{equation}\label{1}
\begin{array}{l}
{T_1} = \left( {\begin{array}{*{20}{c}}
{{e^{ - i\theta }}}&0\\
0&{{e^{i\theta }}}
\end{array}} \right)\left( {\begin{array}{*{20}{c}}
{1 + i\alpha }&{i\alpha }\\
{ - i\alpha }&{1 - i\alpha }
\end{array}} \right)\\\\
 = \left( {\begin{array}{*{20}{c}}
{{e^{ - i\theta }}(1 + i\alpha )}&{{e^{ - i\theta }}i\alpha }\\
{ - {e^{i\theta }}i\alpha }&{{e^{i\theta }}(1 - i\alpha )}
\end{array}} \right)
\end{array}
\end{equation}
where $\alpha=\Gamma/(\omega-\Omega)$, $\Omega$ is the qubit
excitation frequency, $\Gamma$ is the rate of spontaneous emission
of excited qubit to a waveguide. It is proportional to the photon-
qubit interaction.

We assume that $\Omega$ is much larger than the cutoff frequency
of the waveguide, then the dispersion relation of photons at near
resonant frequency, $\omega\approx\Omega$ , can be taken as
linear: $\omega=v_g k$, where $v_g$ is the group velocity.

The definition (\ref{1}) is slightly different from that in
\cite{Shen05} by the inclusion of the phase matrix for the first
qubit:

\begin{equation}\label{2}
\left( {\begin{array}{*{20}{c}}
{{e^{ - i\theta }}}&0\\
0&{{e^{i\theta }}}
\end{array}} \right)
\end{equation}
where $\theta=kL=\omega L/v_g$, $L$ is the distance between
neighbor qubits in multi qubit system. The inclusion of the phase
matrix (\ref{2}) for the first qubit in the array is made solely
for convenience of calculations. It does not influence the
absolute values of reflection or transmission amplitudes.

We define the transfer matrix, $T^{(n)}$ for $n$-th qubit in the
chain in such a way that it relates incoming, $a_{n-1}$ and
outgoing, $b_{n-1}$ wave amplitudes before the qubit to outgoing,
$a_{n}$ and incoming, $b_{n}$ wave amplitudes behind the qubit.

\begin{equation}\label{box}
\begin{array}{*{20}{c}}
  {\xrightarrow{{{a_{n - 1}}}}} \\
  {\xleftarrow{{{b_{n - 1}}}}}
\end{array}\boxed{T_{_{}}^{(n)}}\begin{array}{*{20}{c}}
  {\xrightarrow{{{a_n}}}} \\
  {\xleftarrow{{{b_n}}}}
\end{array}\nonumber
\end{equation}

\begin{equation}\label{t_n}
\left( {\begin{array}{*{20}{c}}
  {{a_{n - 1}}} \\
  {{b_{n - 1}}}
\end{array}} \right) = \left( {\begin{array}{*{20}{c}}
  {T_{_{11}}^{(n)}}&{T_{_{12}}^{(n)}} \\
  {T_{_{21}}^{(n)}}&{T_{_{22}}^{(n)}}
\end{array}} \right)\left( {\begin{array}{*{20}{c}}
  {{a_n}} \\
  {{b_n}}
\end{array}} \right)
\end{equation}
where $a_0=1$, $b_0=r$ is the reflection amplitude.

Transfer matrix for N identical qubits is calculated as the $N$-th
power of the matrix $T_1$:
\begin{equation}\label{3}
    T_N=(T_1)^N
\end{equation}
so that
\begin{equation}\label{t_N}
\left( {\begin{array}{*{20}{c}}
  1 \\
  r
\end{array}} \right) = \left( {\begin{array}{*{20}{c}}
  {{{\left( {{T_N}} \right)}_{11}}}&{{{\left( {{T_N}} \right)}_{12}}} \\
  {{{\left( {{T_N}} \right)}_{21}}}&{{{\left( {{T_N}} \right)}_{22}}}
\end{array}} \right)\left( {\begin{array}{*{20}{c}}
  t \\
  0
\end{array}} \right)
\end{equation}
with $r$ and $t$ being the reflection and transmission amplitudes
for $N$- qubit chain, respectively. Below, we use the engineering
notations for these quantities: $r\equiv S_{11}$, $t\equiv
S_{21}$.

From (\ref{t_N}) the transmission and reflection amplitudes are as
follows:

\begin{equation}\label{S21}
    S_{21}=\frac{1}{(T_N)_{11}}
\end{equation}

\begin{equation}\label{R11}
    S_{11}=\frac{(T_N)_{21}}{(T_N)_{11}}
\end{equation}

Next, we use the Abeles theorem\cite{Abel1950}(see Eq.\ref{Abel}
in appendix), which allows us to write matrix $T_N$ in terms of
Chebyshev polynomials of second kind:
\begin{equation}\label{4}
    T_N=U_{N-1}(y)T_1-U_{N-2}(y)I
\end{equation}
where
\begin{equation}\label{5}
    y = \frac{1}{2}Sp({T_1}) = \cos \theta  + \alpha \sin \theta
\end{equation}

From (\ref{4}) and (\ref{1}) we obtain
\begin{equation}\label{T11}
\begin{array}{l}
{({T_N})_{11}} = {U_{N - 1}}(y){({T_1})_{11}} - {U_{N - 2}}(y)\\\\
 = {U_{N - 1}}(y){e^{ - i\theta }}(1 + i\alpha ) - {U_{N - 2}}(y)
\end{array}
\end{equation}

\begin{equation}\label{Abs1}
\begin{array}{l}
{\left| {{{({T_N})}_{11}}} \right|^2} = U_{_{N - 1}}^2(y)(1 + {\alpha ^2}) + U_{_{N -
2}}^2(y)\\\\
 - 2y{U_{N - 1}}(y){U_{N - 2}}(y) = 1 + {\alpha ^2}U_{_{N - 1}}^2(y)
\end{array}
\end{equation}

\begin{equation}\label{T21}
    (T_N)_{21}=-i\alpha e^{i\theta}U_{N-1}(y)
\end{equation}

On deriving the expression (\ref{Abs1}) we use the definition
(\ref{5}) of $y$ and the identity (\ref{A8}) from appendix.

Therefore, the absolute values of the transmission and reflection
amplitudes can be written in the following form:
\begin{equation}\label{Abs2}
    {\left| {{S_{21}}} \right|^2} = \frac{1}{{1 + {\alpha ^2}{U_{N - 1}^2}(y)}}
\end{equation}

\begin{equation}\label{S11}
{\left| {{S_{11}}} \right|^2} = \frac{{{\alpha ^2}U_{_{N -
1}}^2(y)}}{{1 + {\alpha ^2}U_{_{N - 1}}^2(y)}}
\end{equation}

The energy conservation clearly follows from (\ref{Abs2}) and
(\ref{S11})

\begin{equation}\label{en}
{\left| {{S_{11}}} \right|^2} + {\left| {{S_{21}}} \right|^2} = 1
\end{equation}

If the distance between neighbor qubits is small compared to the
wavelength ($\theta\ll 1$), then $y=1, U_{N-1}(1)=N$ and we obtain
for $S_{21}$:
\begin{equation}\label{Abs21}
{\left| {{S_{21}}} \right|^2} = \frac{{{{(\omega  - \Omega
)}^2}}}{{{{(\omega  - \Omega )}^2} + {{(N\Gamma )}^2}}}
\end{equation}

This is well known phenomena of a superradiance: closely spaced
$N$ identical qubits decay $N$ times faster than a single qubit
\cite{Dicke54}.

From the definition of $U_n(y)$ (see (\ref{A2}) in appendix) it
follows that for $|y|>1$ the Chebyshev polynomial $U_n(y)$ can
take sufficiently large values which give rise to a strong
suppression of the transmission.

The formation of the bandgap begins when the number of qubits $N$
exceeds the value $\sqrt{\Omega/\Gamma}$ \cite{Chang2012}. For
real atoms in nanofiber this value is on the order of $10^4$
\cite{Chang2012}. However, for solid state qubits in the regime of
strong coupling ($\Gamma/\Omega\approx 0.1$) the quantity
$\sqrt{\Omega/\Gamma}\approx 3$. Therefore, for this system the
formation of the bandgap starts with several qubits in the chain.

\begin{figure}
  \includegraphics[width=8 cm]{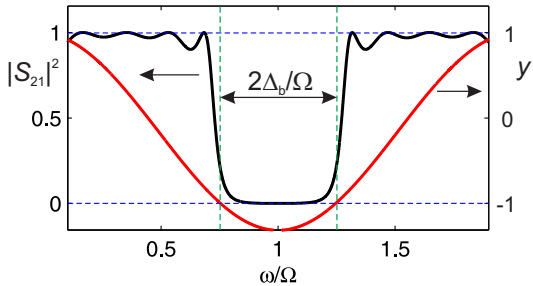}\\
  \caption{Transmission amplitude $|S_{21}|^2$ for five identical qubits (black,
  left Y-axis) and $y$ (red, right Y-axis) vs frequency
   for $k_0L=\pi$, $\Gamma/\Omega=0.1$.}
  \label{fig1}
\end{figure}

The plots of transmission amplitude (\ref{Abs2}) for five
identical qubits together with the frequency dependence of $y$ are
shown in Fig.\ref{fig1}. All qubits are assumed to satisfy the
Bragg condition: $\theta_0=k_0L=\frac{\Omega}{v_g}L=\pi$. We see
that in vicinity of $\theta\approx\pi$ there exists a broad
bandgap where the transmission is strongly suppressed, and within
the band $y<-1$.

For subsequent calculations it is convenient to write $\theta$ as

\begin{equation}\label{theta}
    \theta=\frac{\omega}{v_g}L=k_0L\left(1+\frac{\Delta}{\Omega}\right)
\end{equation}
where $k_0=\Omega/v_g$, $\Delta=\omega-\Omega$.

We take $k_0L=\pi$ and assume $\Delta\ll\Omega$. Thus we obtain for
$y$

\begin{equation}\label{y}
y =  - \cos \left( {\frac{{\pi \Delta }}{\Omega }} \right) -
\frac{\Gamma }{\Delta }\sin \left( {\frac{{\pi \Delta }}{\Omega }}
\right) \approx b + a{\left( {\frac{\Delta }{\Omega }} \right)^2}
\end{equation}
where
\begin{equation}\label{ab}
b =  - 1 - \frac{{\Gamma \pi }}{\Omega };\quad a = \frac{{{\pi
^2}}}{2}\left( {1 + \frac{{\Gamma \pi }}{{3\Omega }}} \right)
\end{equation}

 Next, we assume that the walls of the band correspond to
$y=y_b=-1$. Equating (\ref{y}) to the boundary value $y_b=-1$ we
obtain for half bandwidth
\begin{equation}\label{bw}
\frac{\Delta_b }{\Omega } = \frac{1}{\pi }\sqrt {\frac{{2G\pi
}}{{1 + \frac{{G\pi }}{3}}}}
\end{equation}
where $G=\Gamma/\Omega$, $\Delta_b=\omega_b-\Omega$.

\begin{figure}
  \includegraphics[width=8cm]{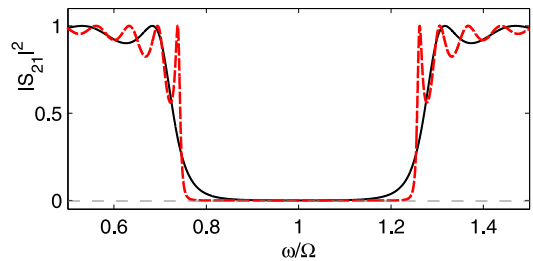}\\
  \caption{Comparison of two bandgaps,
  for $N=5$ (black, solid line) and $N=11$ (red, dashed line).}\label{bgp}
\end{figure}

It should be noted that the estimation (\ref{bw}) does not depend
on the number of qubits. However, a weak dependence of the bandgap
width on $N$ does exist, while the steepness of the bandgap walls
depends on $N$ much more stronger. This is seen in Fig.\ref{bgp}
where we compare two bandgaps, for $N=5$ and $N=11$.

Next we estimate the value of transmission amplitude at the
boundary of the band, where $y\approx -1$ ,
$U_{N-1}(-1)=(-1)^{N-1}N$ (\ref{A14}).

At the boundary
\begin{equation}\label{alfa}
\alpha_b  = \frac{\Gamma }{\Delta_b } = \frac{\Gamma }{\Omega
}\frac{\Omega }{\Delta_b} \approx G\pi \sqrt {\frac{{1 +
\frac{{G\pi }}{3}}}{{2G\pi }}}
\end{equation}

The substitution of (\ref{alfa}) in (\ref{Abs2}) leads to the
following estimation for the amplitude of transmission at the
boundary of the band:
\begin{equation}\label{tr}
\left| {{S_{21}}} \right|_b^2 = \frac{1}{{1 + {N^2}\frac{{G\pi
}}{2}\left( {1 + \frac{{G\pi }}{3}} \right)}}
\end{equation}

\begin{figure}
  \includegraphics[width=8 cm]{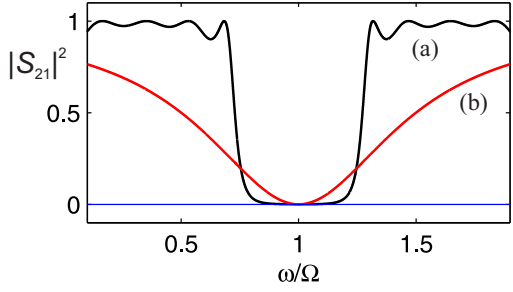}\\
  \caption{Transmission amplitude $|S_{21}|^2$ vs frequency for five identical
  qubits, $\Gamma/\Omega=0.1$.
  (a)(black), calculated from expression (\ref{Abs2}) for $k_0L=\pi$,
  (b)(red) calculated from expression (\ref{Abs21}) which is obtained from (\ref{Abs2})
   for $kL=\pi$.}\label{fig2}
\end{figure}

It is worthwhile to note that if  we disregarded the frequency
dependence of $\theta$ and simply put $\theta=n\pi$, where $n$ is
integer (as was done in \cite{Mukh2019}), the essential
interference effects which are contained in the second term in
(\ref{5}) would be lost. In this case we would obtain a broadened
Lorentzian (\ref{Abs21}) for the transmission amplitude. As the
example, we compare in Fig.\ref{fig2} the transmission amplitudes
for five identical qubits calculated from exact expression
(\ref{Abs2}), where the frequency dependence of the phase $\theta$
is accounted for, with the one calculated from (\ref{Abs21}),
where the frequency dependence of the phase is ignored. The
bandgap (black curve is obtained for $k_0L=\pi$, whereas the broad
Lorentzian- type (red) curve is obtained for $kL=\pi$, where the
frequency dependence of the phase is disregarded.

\begin{figure}
  \includegraphics[width=8.5cm]{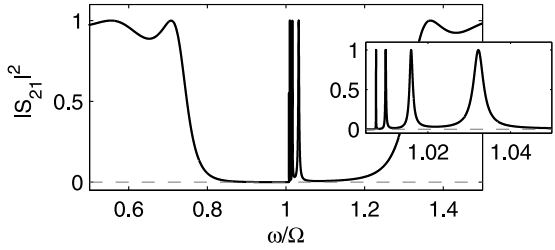}\\
  \caption{Transmission amplitude $|S_{21}|^2$ vs frequency for five identical
  qubits, $\Gamma/\Omega=0.1$, calculated from expression (\ref{Abs2}) for
  $k_0L=0.95\pi$. The fine structure of four peaks is shown in the insert.}\label{Fig10}
\end{figure}

It is known that for $N$- qubit chain there are $N-1$ transmission
peaks, where the reflection is zero \cite{Tsoi2008}. This property
follows directly from (\ref{S11}): the Chebyshev polynomial
$U_{N-1}(y)$ has exactly $N-1$ roots (see (\ref{Root})). The
shallow peaks of the wavy structure which are seen in
Fig.\ref{fig1} before and after the bandgap correspond to four
roots of $U_4$(y), where the reflection amplitude (\ref{S11})
equals zero. However, if all qubits simultaneously are slightly
detuned from exact Bragg resonance ($\Omega/v_g=\xi n\pi$, where
$|1-\xi|<<1$), then $N-1$ transmission peaks fall within the
bandgap. This is shown for five qubits in Fig.\ref{Fig10} for
$\xi=0.95$. All peaks are grouped near $\Omega$. For periodic
structure with infinite $N$ all peaks merge into a single one in
the point $\omega=\Omega$ \cite{Deych2000}.

\section{Tunable resonance within the bandgap}
Here we consider the chain with odd number of qubits, $N=2n+1$,
where $n$ is a positive integer. All qibits in the chain are
supposed to be identical except for the central qubit whose
excitation frequency $\Omega_0$ differs from that of the other
qubits in the chain. We show that in this case there exists within
a stopband a narrow resonance which provides a full transmission.
The position and the width of the resonance depend on the ratio
$\Omega_0/\Omega$ of the frequency of the central qubit,
$\Omega_0$ to the frequency of other qubits, $\Omega$.

\subsection{Basic expressions}

The transfer matrix for the qubit structure described above is of
the form:
\begin{equation}\label{Tn}
{T_{2n+1}} = T_1^n{T_0}T_1^n
\end{equation}
where $T_1$ is given in (\ref{1}), and

\begin{equation}\label{T0}
    \begin{array}{l}
{T_0} = \left( {\begin{array}{*{20}{c}}
{{e^{ - i\theta }}}&0\\
0&{{e^{i\theta }}}
\end{array}} \right)\left( {\begin{array}{*{20}{c}}
{1 + i\alpha_0 }&{i\alpha_0 }\\
{ - i\alpha_0 }&{1 - i\alpha_0 }
\end{array}} \right)\\\\
 = \left( {\begin{array}{*{20}{c}}
{{e^{ - i\theta }}(1 + i\alpha_0 )}&{{e^{ - i\theta }}i\alpha_0 }\\
{ - {e^{i\theta }}i\alpha_0 }&{{e^{i\theta }}(1 - i\alpha_0 )}
\end{array}} \right)
\end{array}
\end{equation}
where $\alpha_0=\Gamma/(\omega-\Omega_0)$, $\Omega_0$ is the
excitation frequency of the central qubit.

The application of Abeles theorem to the matrix $T_1^n$ gives
\begin{equation}\label{Tn1}
\begin{array}{l}
{T_{2n+1}} = U_{n - 1}^2(y){T_1}{T_0}{T_1} - {U_{n - 1}}(y){U_{n -
2}}(y){\left[ {{T_1}{T_0}} \right]_ +
}\\\\
 + U_{n - 2}^2(y){T_0}
\end{array}
\end{equation}
where
\[[T_0T_1]_+=T_0T_1+T_1T_0\]
Next we define the quantity which characterizes the difference
between the central and other qubits in the chain:
\begin{equation}\label{d}
    \delta=\alpha_0-\alpha=\alpha_0\alpha\eta
\end{equation}
where $\eta=(\Omega_0-\Omega)/\Gamma)$.

Therefore, the matrix $T_0$ can be written as
\begin{equation}\label{T00}
    T_0=T_1+i\delta R
\end{equation}
where
\begin{equation}\label{R}
R = \left( {\begin{array}{*{20}{c}}
{{e^{ - i\theta }}}&{{e^{ - i\theta }}}\\
{ - {e^{i\theta }}}&{ - {e^{i\theta }}}
\end{array}} \right)
\end{equation}

If $\delta=0$ the expression (\ref{Tn1}) reduces to (\ref{4}).
Therefore, we can rewrite (\ref{Tn1}) as follows:

\begin{equation}\label{TNd}
{T_{2n+1}} = T_N + i\delta X
\end{equation}
where $T_N$ is given in (\ref{4}) and
\begin{equation}\label{X}
X = U_{n - 1}^2(y){T_1}R{T_1} - {U_{n - 1}}(y){U_{n -
2}}(y){\left[ {{T_1}R} \right]_ + } + U_{n - 2}^2(y)R
\end{equation}
In this case, the transmission amplitude can be written as follows
\begin{equation}\label{ta}
{\left| {{S_{21}}} \right|^2} = \frac{1}{{{{\left| {{{\left(
{{T_N}} \right)}_{11}} + i\delta {{\left( X \right)}_{11}}}
\right|}^2}}}
\end{equation}

In order to transform (\ref{X}) in more convenient form we
introduce commutator of two matrices, $R$ and $T_1$:
\begin{equation}\label{com}
\left[ {R,{T_1}} \right]_- = R{T_1} - {T_1}R = 2i{e^{i\theta
}}\sin \theta \left( {\begin{array}{*{20}{c}}
0&{{e^{ - i\theta }}}\\
{{e^{i\theta }}}&0
\end{array}} \right)
\end{equation}
and rewrite (\ref{X}) as follows:
\begin{equation}\label{X1}
\begin{array}{l}
X = \left[ {2yU_{n - 1}^2 - 2{U_{n - 1}}{U_{n - 2}}} \right]{T_1}R + U_{n - 1}^2{T_1}{\left[ {R{T_1}} \right]_ -
}\\\\
 - {U_{n - 1}}{U_{n - 2}}{\left[ {R{T_1}} \right]_ - } - \left( {U_{n - 1}^2 - U_{n - 2}^2} \right)R
\end{array}
\end{equation}
With the aid of recurrence relation (\ref{A1}) and properties of
Chebyshev polynomials (\ref{A5}), (\ref{A6}) we write (\ref{X}) in
the final form;
\begin{equation}\label{X2}
\begin{array}{l}
X = {U_{N - 2}}{T_1}R + U_{n - 1}^2{T_1}{\left[ {R{T_1}} \right]_ -
}\\\\
 - {U_{n - 1}}{U_{n - 2}}{\left[ {R{T_1}} \right]_ - } - {U_{N - 3}}R
\end{array}
\end{equation}

\subsection{Frequency of the resonance inside the bandgap}
In what follows we need only two matrix elements $X_{11}$ and
$X_{21}$.
\begin{equation}\label{X11}
\begin{array}{l}
{X_{11}} = {U_{N - 2}}\left[ {{e^{ - 2i\theta }}(1 + i\alpha ) - i\alpha }
\right]\\\\
 - U_{n - 1}^22\alpha {e^{i\theta }}\sin \theta  - {U_{N - 3}}{e^{ - i\theta }}
\end{array}
\end{equation}

\begin{equation}\label{X21}
\begin{array}{l}
{X_{21}} = {U_{N - 2}}\left[ { - i\alpha  - {e^{2i\theta }}(1 - i\alpha )}
\right]\\\\
 + U_{n - 1}^22i{e^{i\theta }}\sin \theta {e^{2i\theta }}(1 - i\alpha
 )\\\\
 - {U_{n - 1}}{U_{n - 2}}2i{e^{2i\theta }}\sin \theta  + {U_{N - 3}}{e^{i\theta }}
\end{array}
\end{equation}
We show below that within a bandgap there exists a frequency where
the reflection amplitude is exactly equal to zero.

For the numerator of reflection amplitude (\ref{R11}) we have
\begin{equation}\label{Num}
   (T_{2n+1})_{21}= -i\alpha e^{i\theta}U_{N-1}(y)+i\delta (X)_{21}
\end{equation}

Inside the bandgap $\theta\approx \pi$, $\alpha\gg 1$. In these
approximations the quantities $X_{11}$ and  $X_{21}$ can be
approximated as

\begin{equation}\label{XX}
{X_{11}} = {U_{N - 2}} + {U_{N - 3}};{\rm{  }}{X_{21}} =  - \left(
{{U_{N - 2}} + {U_{N - 3}}} \right)
\end{equation}
For (\ref{Num}) within the bandgap we thus obtain
\begin{equation}\label{Num1}
     (T_{2n+1})_{21}\approx i\alpha U_{N-1}(y)-i\delta (U_{N - 2} + U_{N -
     3})
\end{equation}
Making use of (\ref{d}) we can obtain from (\ref{Num1}) the
equation which defines the frequency of the resonance within the
bandgap
\begin{equation}\label{Rp}
 U_{N-1}(y)-\alpha_0 \eta(U_{N - 2}(y) + U_{N - 3}(y))=0
\end{equation}
From (\ref{Rp}) we obtain the expression for the resonance
frequency, $\omega_p$  which is valid for $\Delta\ll \Omega$.
\begin{equation}\label{omp}
{\omega _p} = {\Omega _0} + \left( {{\Omega _0} - \Omega }
\right)\frac{{{U_{N - 2}}({b}) + {U_{N - 3}}({b})}}{{{U_{N -
1}}({b})}}
\end{equation}
where $b$ is defined in (\ref{ab}).

From (\ref{omp}) we see that the resonance frequency moves to the
walls of the band as the frequency of the central qubit deviates
from $\Omega$. The corresponding picture is shown in
Fig.\ref{Fig3} for five qubits with $\Gamma/\Omega=0.1$.

Two dimensional distribution of transmission amplitude for five
qubits against the frequency, $\omega$ and the ratio
$\Omega_3/\Omega$ is shown in Fig.\ref{Fig3}. The resonance line
is seen as a narrow colored stripline  in the middle of the
bandgap.

\begin{figure}
  \includegraphics[width=8cm]{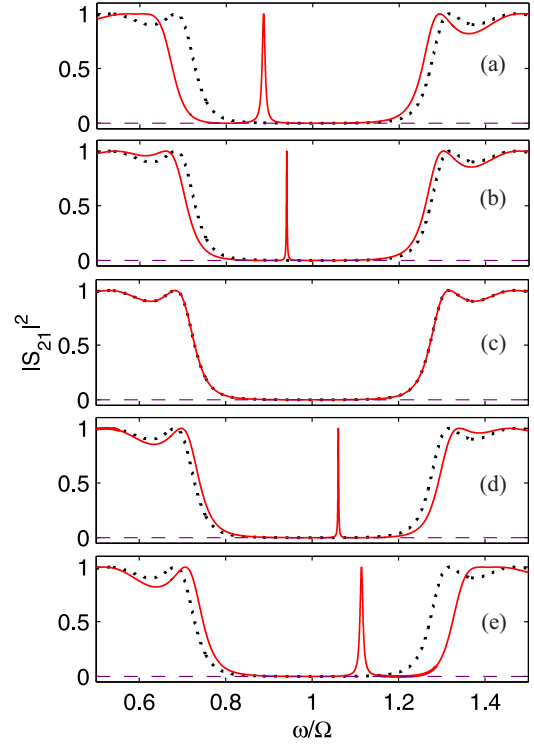}\\
  \caption{Movement of resonance line inside the bandgap as the ratio
   $\Omega_0/\Omega$ is increased. $N=5$, $\Gamma/\Omega=0.1$. (a) $\Omega_0/\Omega=0.8$,
   (b) $\Omega_0/\Omega=0.9$,(c) $\Omega_0/\Omega=1.0$, (d) $\Omega_0/\Omega=1.1$,
   (e) $\Omega_0/\Omega=1.2$. Solid (red) line is for the chain with a central
    qubit being different. Dashed (black) line is for the chain
    where all qubits are identical.
   }\label{Fig6}
\end{figure}

\begin{figure}
  \includegraphics[width=8 cm]{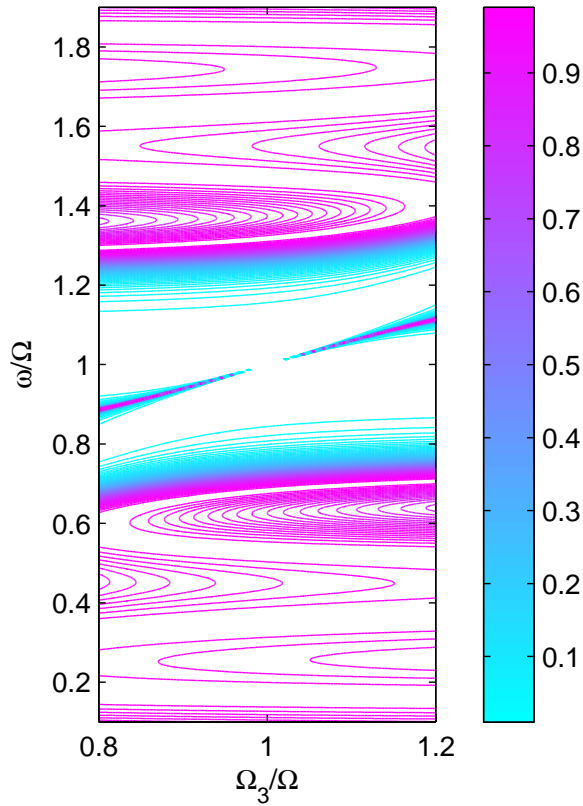}\\
  \caption{2D spatial distribution of the transmission amplitude $|S_{21}|^2$
  for five qubit chain where $\Omega_3$ is the excitation
  frequency of a third qubit. The color scale gives the amplitude $|S_{21}|^2$
  of transmitted signal.}\label{Fig3}
\end{figure}

\subsection{The lineshape and the width of the resonance}
In order to derive resonance lineshape we need to calculate the
quantity $|(T_{2n+1})_{11}|^2$ which defines the transmission
amplitude $|S_{21}|^2$.
\begin{equation}\label{L1}
\begin{array}{l}
{\left| {{{\left( {{T_{2n + 1}}} \right)}_{11}}} \right|^2} = {\left| {{{\left( {{T_N}} \right)}_{11}} + i\delta {{\left( X \right)}_{11}}}
\right|^2}\\\\
 = {\left| {{{\left( {{T_N}} \right)}_{11}}} \right|^2}
 + {\delta ^2}{\left| {\left( X \right)_{11}^{}} \right|^2}\\\\
  + i\delta \left[ {{{\left( X \right)}_{11}}\left( {{T_N}} \right)_{_{11}}^* - \left( X \right)_{11}^*\left( {{T_N}} \right)_{_{11}}^{}} \right]
\end{array}
\end{equation}
Making use of (\ref{Abs1}) and the approximation (\ref{XX}) for
$(X)_{11}$, we obtain from (\ref{L1}):
\begin{equation}\label{L2}
{\left| {{{\left( {{T_{2n + 1}}} \right)}_{11}}} \right|^2} = 1 +
{\alpha ^2}{\left[ {{U_{n - 1}} - {\alpha _0}\eta \left( {{U_{N -
2}} + {U_{N - 3}}} \right)} \right]^2}
\end{equation}
The quantity in square brackets in the right hand side of
(\ref{L2}) is just the one in the left hand side of (\ref{Rp}),
which defines the position of resonance inside the band. As is
seen from (\ref{L2}) at the point of resonance a full transmission
is observed.

The investigation of transmission amplitude $S_{21}$ near
resonance leads to the following expression for the spectrum of
transmission line (the details of derivation are given in the
appendix):
\begin{equation}\label{L3}
{S_{21}}(\omega ) = \frac{{\omega  - \Omega }}{{\omega  - \Omega +
i\Gamma {B_N}({y_p})(\omega  - {\omega _p})}}
\end{equation}
where the quantity $B_N(y_p)$ is given in (\ref{B9}):
\begin{equation}\label{BN}
\begin{array} {l}
{B_N}({y_p}) = \frac{{2a{\Delta _p}}}{{{\Omega ^2}}}\left( {U{'_{N
- 1}} - {\alpha _0}({\omega _p})\eta \left[ {U{'_{N - 2}} + U{'_{N
- 3}}} \right]} \right)\\\\ + \frac{{{U_{N - 1}}}}{{{\Delta
_{p,0}}}}
\end{array}
\end{equation}

\begin{figure}
  \includegraphics[width=8 cm]{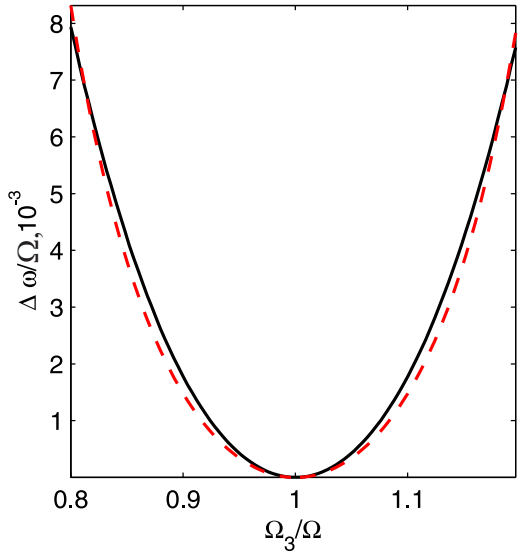}\\
  \caption{The width of the resonance line against the frequency of a central qubit. $N=5$, $G=0.1$.
  Solid (black) plot is calculated from exact expression (\ref{ta}), dashed (red) plot
  is calculated from the estimation (\ref{L6}).}\label{width}
\end{figure}

\begin{figure}
  \includegraphics[width=8cm]{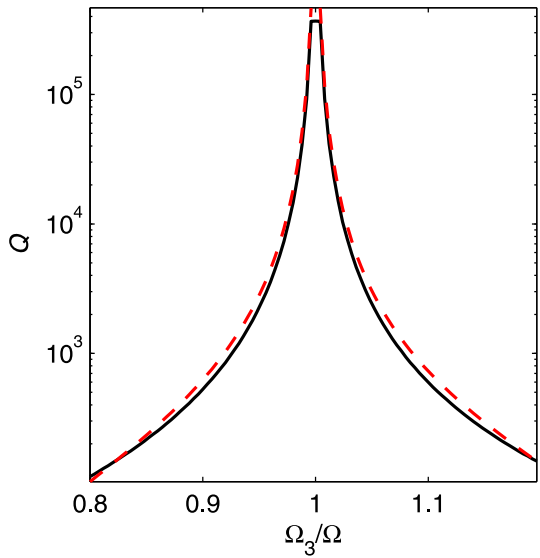}\\
  \caption{The quality factor of the resonance line against the frequency of a central qubit. $N=5$, $G=0.1$.
  Solid (black) plot is calculated from exact expression (\ref{ta}), dashed (red) plot
  is calculated from the estimation (\ref{L6})}\label{qual}
\end{figure}

It is seen from expression (\ref{L3}) that the lineshape of
resonance inside the band is not the Lorentzian function: its
width depends on the frequency. The full width at half maximum
(FWHM) of the resonance we find from the equation
\begin{equation}\label{L4}
    \mid S_{21}(\omega_{\pm})\mid^2=\frac{1}{2}
\end{equation}
We thus obtain:
\begin{equation}\label{L5}
{\omega _ + } = \frac{{\Omega  - \Gamma {B_N}{\omega _p}}}{{1 -
\Gamma {B_N}}};\;{\omega _ - } = \frac{{\Omega  + \Gamma
{B_N}{\omega _p}}}{{1 + \Gamma {B_N}}}
\end{equation}
Finally, the FWHM of the resonance line is as follows:
\begin{equation}\label{L6}
\Delta \omega  = {\omega _ + } - {\omega _ - } = \frac{{2\Gamma
{B_N}(\Omega  - {\omega _p})}}{{1 - {\Gamma ^2}B_{^N}^2}}
\end{equation}
It follows from (\ref{L6}) that the width of resonance line
depends on its position within the bandgap. The closer the
resonance line to the wall of the bandgap, the greater its width.
The plots of the width of the resonance line and the corresponding
quality factor, $Q=\omega_p/\Delta\omega$  against the frequency
of central qubit are shown for five-qubit structure in
Fig.\ref{width} and Fig.\ref{qual}, respectively. The solid
(black) curves are calculated from exact expression (\ref{ta}),
the dashed (red) curves are calculated from the estimation
(\ref{L6}).

\begin{figure}
  \includegraphics[width=8cm]{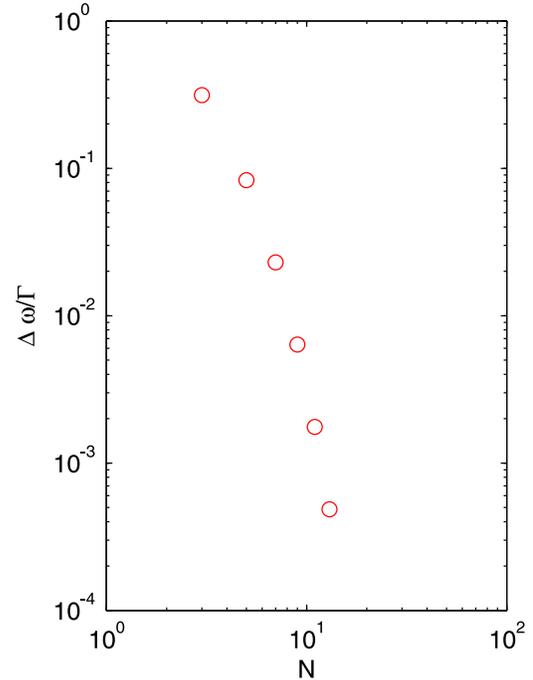}\\
  \caption{The dependence of resonance width on the number of qubits in a chain.
  The calculation is performed for first $N$- odd chains, $N=3, 5, 7, 9, 11, 13$.
  $\Omega_0/\Omega=1.2$, $\Gamma/\Omega=0.1$.}\label{DwN}
\end{figure}

 In fact, the resonance line inside the bandgap
corresponds to a subradiant state. Its decay rate is just the
width of resonance line. In Fig.\ref{DwN} we compare the decay
rate for these subradiant states with that of individual qubits
for several $N$- odd qubit chains. We see from Fig.\ref{DwN} that
as the number of qubits is increased, the resonance width is
strongly reduced.

\begin{figure}
  \includegraphics[width=8 cm]{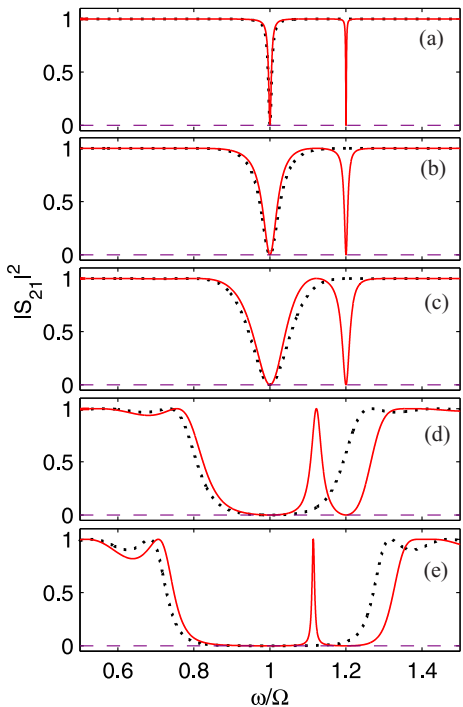}\\
  \caption{The process of the peak formation for five- qubit system,
  $\Omega_3/\Omega=1.2$, (a) $G=0.001$, (b) $G=0.005$, (c) $G=0.01$,
   (d) $G=0.05$, (e) $G=0.1$.}\label{form}
\end{figure}

It is interesting to look at the process of a formation of the
bandgap and that of the resonance line. Consider the dependence of
the transmission curve on the rate of single qubit spontaneous
emission, $\Gamma$ into a waveguide. This rate is the measure of
the qubit- photon interaction. If all qubits are identical and the
qubit- photon interaction is weak, the transmission amplitude
looks like the one shown by dashed (black) curve in
Fig.\ref{form}, panel (a).  As the qubit- photon interaction is
increased, the transmission curve starts to deform, the width of
the band becomes larger, and finally it transforms to the
structure with a bandgap as is shown in the panel (e) of
Fig.\ref{form}.  This is a direct manifestation of the formation
of a superradiant state which results from the overlapping of the
qubit individual widths \cite{Auerbach2011}.

If one qubit has a different frequency, $\Omega_0$, and the qubit-
photon interaction is weak, then the transmission curve represents
two dips, at $\Omega$ and $\Omega_0$ as is shown by solid (red)
curve in the panel (a) in Fig.\ref{form}.   Between these two dips
there is a broad region where transmission equals unity. As the
qubit- photon interaction is increased, the region between two
dips (panel (a)) undergoes transformation which finally results in
a single narrow resonance where the transmission is equal to unity
(panel (e) in Fig.\ref{form}). The reason for this transformation
is the same as above: the photon-qubit interaction leads to the
overlapping of the qubit individual widths.

\begin{figure}
  \includegraphics[width=8 cm]{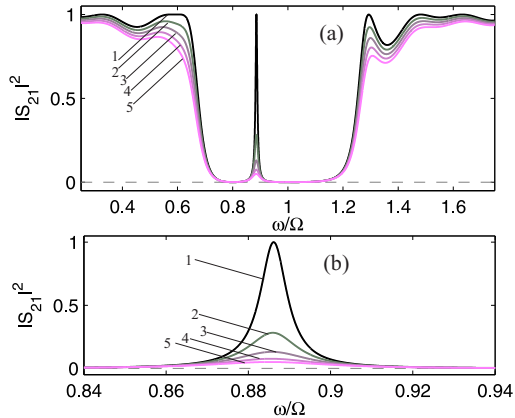}\\
  \caption{(a) Frequency dependence of the photon transmission through
  five-qubit chain for different values of the radiative losses.
  (b) Lineshape of the transparency peak
  inside the band-gap for different values of the radiative losses.
  $k_0L=\pi$, $\Omega_{1,2,4,5}=\Omega-i\gamma$, $\Omega_3=0.8\Omega-i\gamma$,
  $\Gamma/\Omega=0.1$, (1)$\gamma/\Omega=0$, (2) $\gamma/\Omega=0.005$,
  (3) $\gamma/\Omega=0.01$, (4) $\gamma/\Omega=0.015$, (5) $\gamma/\Omega=0.02$.}
  \label{fig11}
\end{figure}

Finally, we consider how the transmission is influenced by the
radiative losses to the environment (out of the waveguide). It is
common to model these losses by the modification of the qubit
frequency with a small shift to lower half of the complex plane:
$\Omega\rightarrow \Omega-i\gamma$, where $\gamma$ is the
radiation rate to the environment. The influence of radiative
losses is shown in Fig.\ref{fig11}. The calculation is made for
five-qubit chain, for $k_0L=\pi$, $\Gamma/\Omega=0.1$, and five
gradually increasing values of $\gamma$. We see (the panel (a) in
Fig.\ref{fig11})) that the width  of the band-gap near its bottom
and the position of transmission peak are practically not
sensitive to the $\gamma$. However, the lineshape of the
transmission peak (the panel (b) in Fig.\ref{fig11})) changes
drastically: as $\gamma$ increases the height of the peak
decreases, and its width increases.

\section{Conclusion}

We have analyzed the signal transmission through a 1D chain of
equally spaced $N$ qubits using the decomposition of a transfer
matrix in terms of Chebyshev polynomials. We find that if the
phase factor at the qubit frequency is equal to integer multiple
of $\pi$, the bandgap is developed in the spectrum of transmitted
signal.

Detailed analysis have been performed for the structure where
$N-1$ qubits were identical in their frequency, $\Omega$, but the
frequency of a central qubit, $\Omega_0$ was different. In this
case, a narrow resonance line appeared within the bandgap with its
position and the width being dependent on the ratio
$\Omega_0/\Omega$.

We showed that the the appearance of the bandgap and the narrow
resonance resulted from the strong photon- qubit interaction when
the widths of individual qubits were overlapped.

Our results may be applied not only to superconducting qubits
\cite{Brehm2020, Mir2019}, but to many other quantum emitters,
interacting with any kind of quasi-1D photonic structures or
circuits.

\begin{acknowledgments}
Ya. S. G. thanks A. Sultanov for fruitful discussions. The work is
supported by the Ministry of Education and Science of Russian
Federation under the project FSUN-2020-0004.
\end{acknowledgments}

\appendix

\section{Basic properties of Chebyshev polynomials of second kind, $U_n(y)$}

1. Definition
\begin{equation}\label{A2}
{U_{N - 1}}(y) = \left\{ {\begin{array}{*{20}{c}}
\begin{array}{l}
\frac{{\sin N\Lambda }}{{\sin \Lambda }},{\kern 1pt} \quad \Lambda
 = \arccos (y),{\kern 1pt} \quad |y| \le 1\\\\
\end{array}\\
\begin{array}{l}
{(sign(y))^{N - 1}}\frac{{\sinh N\Lambda }}{{\sinh \Lambda }},{\kern 1pt}
\\\\
\quad \Lambda  = \ln (|y| + \sqrt {|y{|^2} - 1} ),{\kern 1pt}
\quad |y| > 1
\end{array}
\end{array}} \right.
\end{equation}

2. Recurrence relation
\begin{equation}\label{A1}
    {U_N} = 2y{U_{N - 1}} - {U_{N - 2}}
\end{equation}

3. Roots

Chebyshev polynomial $U_{N-1}(y)$ has $N-1$ roots $y_m$ where
\begin{equation}\label{Root}
    y_m=\cos\left(\frac{\pi m}{N}\right)
\end{equation}
with $m=1, 2,.....N-1$

4. Abeles theorem \cite{Abel1950}.

The $n$-th power $M^n$ of any unimodular (determinant is equal to
1) matrix $M$ can be expressed in terms of Chebyshev polynomials
of second kind as follows:
\begin{equation}\label{Abel}
    M^n = {U_{n - 1}}(y)M - {U_{n - 2}}(y)I
\end{equation}
where $y=\frac{1}{2}Sp(M)$ and $I$ is the identity matrix.

5. Identities
\begin{equation}\label{A3}
    {U_{2n + 1}} = 2{U_n}(y{U_n} - {U_{n - 1}})
\end{equation}

\begin{equation}\label{A4}
    {U_{2n - 1}} = 2{U_{n - 1}}(y{U_{n - 1}} - {U_{n - 2}})
\end{equation}

\begin{equation}\label{A5}
    {U_{n + m}} = {U_n}{U_m} - {U_{n - 1}}{U_{m - 1}}
\end{equation}

\begin{equation}\label{A6}
    {U_{n - 1}}({U_n} - {U_{n - 2}}) = {U_{2n - 1}}
\end{equation}

\begin{equation}\label{A8}
U_{N - 1}^2(y) + U_{N - 2}^2(y) - 2y{U_{N - 1}}(y){U_{N - 2}}(y) =
1
\end{equation}

6. First several polynomials
\begin{equation}\label{A9}
    U_{-1}(y)=0;\quad U_0(y)=1
\end{equation}

\begin{equation}\label{A10}
    U_1(y)=2y
\end{equation}

\begin{equation}\label{A11}
    U_2(y)=4y^2-1
\end{equation}

\begin{equation}\label{A12}
    U_3(y)=8y^3-4y
\end{equation}

\begin{equation}\label{A13}
U_4(y)=16y^4-12y^2+1
\end{equation}

7. Specific values
\begin{equation}\label{A14}
    U_n(\pm 1)=(\pm 1)^n(n+1)
\end{equation}

\begin{equation}\label{A15}
{U_{2n}}(0) = {( - 1)^n};\;\quad {U_{2n + 1}}(0) = 0
\end{equation}

\section{Derivation of the resonance lineshape}
The expression (\ref{L2}) allows us to write the transmission
amplitude in the following form
\begin{equation}\label{B1}
{S_{21}}(y) = \frac{1}{{1 + i\alpha {A_N}(y)}}
\end{equation}
where
\begin{equation}\label{B2}
{A_N}(y) = {U_{N - 1}}(y) - {\alpha _0}(\omega )\eta \left( {{U_{N
- 2}}(y) + {U_{N - 3}}(y)} \right),
\end{equation}
where
\begin{equation}\label{B3}
{\alpha _0}(\omega ) = \frac{\Gamma }{{{\omega _p} - \Omega_0  +
\delta \omega }} \approx {\alpha _0}({\omega _p}) - {\alpha
_0}({\omega _p})\frac{{\delta \omega }}{\Delta_{p,0}},
\end{equation}
$\alpha_0(\omega_p)=\Gamma/\Delta_{p,0}$,
$\Delta_{p,0}=\omega_p-\Omega_0$, $\delta\omega=\omega-\omega_p$.

Let at the point of resonance where $A_N(y)=0$
\begin{equation}\label{B4}
y \equiv {y_p} = b + a{\left( {\frac{{{\Delta _p}}}{\Omega }}
\right)^2}
\end{equation}

Near resonance the transmission amplitude can be written as
follows:
\begin{equation}\label{B5}
{S_{21}}({y_p} + \delta y) = \frac{1}{{1 + i\alpha {A_N}({y_p} +
\delta y)}}
\end{equation}
where
\begin{equation}\label{B6}
\delta y = y - {y_p} \approx \frac{{2a{\Delta _p}}}{{{\Omega
^2}}}\delta \omega
\end{equation}
with $\delta\omega=\omega-\omega_p$.

Then, we obtain
\begin{equation}\label{B7}
{A_N}({y_p} + \delta y) \approx {B_N}({y_p})\delta \omega
\end{equation}
where
\begin{equation}\label{B8}
\begin{array}{l}
{B_N}({y_p}) = \frac{{2a{\Delta _p}}}{{{\Omega ^2}}}\left( {U{'_{N - 1}} - {\alpha _0}({\omega _p})\eta \left[ {U{'_{N - 2}} + U{'_{N - 3}}} \right]}
\right)\\\\
 + \frac{{{\alpha _0}({\omega _p})}}{{{\Delta _{p,0}}}}\eta \left( {{U_{N - 2}} + {U_{N - 3}}} \right)
\end{array}
\end{equation}
In expression (\ref{B8}) the Chebyshev polynomials and their
derivatives are taken at the point $y=y_p$.

The equation (\ref{Rp}) allows us to rewrite  the last term in
(\ref{B8}), so that we finally obtain for $B_N(y_p)$:
\begin{equation}\label{B9}
\begin{array} {l}
{B_N}({y_p}) = \frac{{2a{\Delta _p}}}{{{\Omega ^2}}}\left( {U{'_{N
- 1}} - {\alpha _0}({\omega _p})\eta \left[ {U{'_{N - 2}} + U{'_{N
- 3}}} \right]} \right)\\\\ + \frac{{{U_{N - 1}}}}{{{\Delta
_{p,0}}}}
\end{array}
\end{equation}

\end{document}